\documentclass[fleqn,10pt]{wlscirep}
\usepackage[utf8]{inputenc}
\usepackage[T1]{fontenc}
\title{Validity of annealed approximation in a high-dimensional system}

\author[1]{Jaegon Um}
\author[2,*]{Hyunsuk Hong}
\author[3]{Hyunggyu Park}
\affil[1]{Department of Physics, Pohang University of Science and Technology, Pohang 37673, Korea}
\affil[2]{Department of Physics and Research Institute of Physics and Chemistry, Jeonbuk National University, Jeonju 54896, Korea }
\affil[*]{hhong@jbnu.ac.kr}
\affil[3]{Quantum Universe Center, Korea Institute for Advanced Study, Seoul 02455, Korea}

\begin{abstract}
This study investigates the suitability of the annealed approximation in high-dimensional systems characterized by dense networks with quenched link disorder, employing models of coupled oscillators.
We demonstrate that dynamic equations governing dense-network systems converge to those of the complete-graph version
in the thermodynamic limit, where link disorder fluctuations vanish entirely. Consequently, the annealed-network systems, where
fluctuations are attenuated, also exhibit the same dynamic behavior in the thermodynamic limit. However, a significant discrepancy arises
in the incoherent (disordered) phase wherein the finite-size behavior becomes critical in determining the steady-state pattern.
To explicitly elucidate this discrepancy, we focus on identical oscillators subject to competitive attractive and repulsive couplings.
In the incoherent phase of dense networks, we observe the manifestation of random irregular states. In contrast, the annealed approximation yields a symmetric (regular) incoherent state where two oppositely coherent clusters of oscillators coexist, accompanied by the vanishing order parameter.
Our findings imply that the annealed approximation should be employed with caution even in dense-network systems, particularly
in the disordered phase.
\end{abstract}
\begin{document}

\flushbottom
\maketitle

\thispagestyle{empty}

\section*{Introduction}

Recently, there has been notable attention given to dynamics of complex systems.
One popular strategy for understanding their connection geometry is through the use of networks composed of nodes and links~\cite{network1, network2}.
Network structure often exhibits quenched link disorder, thereby rendering the system analytically intractable.
In addition to numerical analyses, mean-field approximations have been employed to study the collective properties of complex-network systems.
The ``annealed'' approximation (AA), a frequently employed mean-field method~\cite{network1, network2, hetero, annealed1, annealed2, annealed3, annealed4, annealed5, annealed6, annealed7, annealed8, annealed9, annealed10, annealed11}, characterizes a quenched link as an annealed one with an appropriate linking probability.
This approximation is commonly referred to the heterogeneous mean-field approximation~\cite{network1, network2, hetero} in complex-network
studies, where the linking probability depends solely on the numbers of links (degrees) of the two connecting nodes.

In sparse networks characterized by a finite mean degree, it is well-established that the AA manifests various limitations~\cite{annealed6, annealed7, annealed9, annealed10, invalidity}, as it only captures a portion of network disorder.
For instance, in investigations of Kuramoto-type models of coupled oscillators~\cite{ref:KM, review:KM, flat, bimodal}, previous studies have
observed that systems subject to the AA not only shift the transition point but also occasionally alter the nature of the transition~\cite{annealed10}. On the contrary,
in the context of dense networks with a diverging mean degree~\cite{dense_network}, there exists a widely held belief that
such networks closely approximate a complete graph (all-to-all connections) in the thermodynamic limit. Consequently, it is reasonable
to anticipate that the AA may correctly characterize the collective properties of dense-network systems in general, as
the AA mitigates disorder fluctuations in such systems~\cite{dense_validity}.
It is reminiscent of the adiabatic elimination method commonly employed in quantum optics~\cite{adiabatic_elimination}: According to the adiabatic elimination, the dense-network fluctuations can be regarded as {\it fast modes}, resulting in corrections to the {\it slow-mode} dynamics corresponding to the fully connected case.
Nevertheless, in the incoherent (disordered) phase where the order parameter vanishes,
finite-size effects may exert significant influence on stabilizing steady states, suggesting a breakdown of the adiabatic elimination because of a lack of slow-modes.
Thus, one might imagine a potential
disparity in the incoherent
steady-state patterns for systems on the complete graph (CG), dense network (DN), and annealed network (AN),
attributed to the finite-size effects contingent upon the network structure.

In this study, we explore the extent of similarity between a DN system and its CG counterpart, and assess the validity of
the AA. To achieve this, we employ systems of the Kuramoto-type oscillators known for exhibiting collective properties sensitive to
connectivity (link) disorder, which has led to a critical failure of the AA in sparse-network systems~\cite{annealed10}.
Firstly, we demonstrate that connectivity fluctuations in DNs vanish in the thermodynamic limit, resulting in the order parameter
behavior identical to that of the CG version.
In the case of identical oscillators without frequency disorder, often referred as the Watanabe-Strogatz (WS) model~\cite{ws1, ws2},
we observe that finite-size effects stemming from connectivity disorder are strong
enough to destroy all infinitely many, initial-condition dependent, ``regular'' (symmetric)
steady-state patterns found in the incoherent phase of its CG version~\cite{ws1, ws2}. In DNs,
random steady-state patterns emerge in the end, suggesting that quenched disorder can readily eliminate regularity in incoherent patterns.
Under the influence of the AA, similar incoherent regular patterns appear with more complex symmetries, distinct from both the random patterns observed in DNs and the regular patterns in the CG version. We note that all these regular patterns are vulnerable to frequency disorder, resulting in random incoherent patterns. Consequently, the incoherent patterns are all identical in the CG, DN, and AN with heterogeneous oscillators featuring random natural frequencies.

Secondly, we introduce coupling disorder (competition of attractive and repulsive couplings) to the WS model on the CG
and examine its collective behavior both numerically and by applying the AA. This model can be also viewed as the
WS model on a combined version of two DNs based on two competing couplings.
Intriguingly, we discover a single initial-condition independent symmetric steady-state pattern in the incoherent phase for the system subject to the AA.
In this symmetric pattern, oscillators are sharply divided into two coherent clusters with a phase difference of $\pi$.
The sizes of these two clusters become identical in the thermodynamic limit, resulting in the
vanishing order parameter. Our numerical analysis without the AA reveals that this incoherent symmetric pattern is replaced by the incoherent random pattern. This sharp and simple disparity in incoherent states leads us to conclude that the AA may not accurately represent
the incoherent steady-state pattern even in (generalized) DNs for a wide range of various many-body dynamics.

Finally, we note that temporal networks where connection links undergo temporal changes bear resemblance to the annealed network systems when
the time scale for link connection is sufficiently short. Thus, in these networks which are prevalent in various biological and
social systems, the  regular/symmetric incoherent patterns can be empirically  observed.

\section*{Kuramoto model on dense networks}

We start with a system of $N$ Kuramoto-type oscillators ~\cite{ref:KM} on a complex network. The dynamics of each oscillator
is governed by the equation of motion as
\begin{equation}
\label{eq:eom_network}
\dot{\phi}_i = \omega_i + \frac{J}{\langle k \rangle} \sum_{j=1}^{N} a_{ij} \sin \left( \phi_j - \phi_i \right) \,,
~~~~~i=1,\cdots, N,
\end{equation}
where $\phi_i$ and $\omega_i$ represent the phase angle and natural random frequency of oscillator $i$, respectively.
The element $a_{ij}$ of the adjacency matrix denotes the connectivity between oscillators $i$ and $j$, with
$a_{ij}=1$ indicating a connection and $a_{ij}=0$ otherwise. The parameter $J$ represents the strength of coupling.
For simplicity, the summation is normalized by the mean degree $\langle k \rangle = (1/N) \sum_{i} k_{i}$, where $k_i$($=\sum_j a_{ij}$) denotes the degree (the number of neighbors) of oscillator $i$.

Defining the ``local'' field $h_i$ by
\begin{equation}
\label{eq:hi1}
h_i \equiv \frac{1}{\langle k \rangle } \sum_{j} a_{ij} z_j\,
\end{equation}
with phase factor $z_j=e^{ {\rm i} \phi_j }$, Eq.~(\ref{eq:eom_network}) is now rewritten as
\begin{equation}
\label{eq:eom_network2}
\dot{\phi}_i = \omega_i +
J~{\rm {Im}}\left(h_i z_i^* \right) \, ,
\end{equation}
where
${\rm Im}(X)$ denotes the imaginary part of $X$ and $z_i^*$
the complex conjugate of $z_i$.
We also define the order parameters from averaged $z_i$ and $h_i$, respectively:
\begin{align}
\label{eq:op}
&\langle z\rangle \equiv \frac{1}{N} \sum_{i} z_i\,,\nonumber \\
&\langle h \rangle \equiv \frac{1}{N}\sum_{i} h_i = \frac{1}{N\langle k \rangle} \sum_{j} k_j \, z_j
=\frac{ \langle k z \rangle }{ \langle k \rangle }\,.
\end{align}
Note that $\langle z \rangle$ is the so-called Kuramoto phase order parameter defined in Ref.~\citeonline{ref:KM}.

It is convenient to recast the local field in the form as
\begin{equation}
\label{eq:hi2}
h_i =  \frac{1}{\langle k \rangle } \sum_{j} a_{ij} \left( \langle z \rangle + z_j - \langle z \rangle \right)
=\langle z \rangle \left ( 1 + \delta\tilde k_i \right) + \xi_i\,,
\end{equation}
where $\delta\tilde k_i$ and $\xi_i$ represent the degree fluctuation and the average phase factor ($z_i$) fluctuation over neighbors, respectively,
with
\begin{equation} \label{eq:deltak}
\delta\tilde k_i \equiv \frac{k_i - \langle k \rangle }{\langle k \rangle }
\end{equation}
and a ``noise'' term $\xi_i$ reads
\begin{equation}
\label{eq:xi}
\xi_i =\frac{1}{\langle k \rangle } \sum_{j} a_{ij}  \left(z_j - \langle z \rangle \right)\,.
\end{equation}

In the case of the CG where $a_{ij}=1$ for all pairs, the local field and two order parameters become identical, i.e., $h_i = \langle h \rangle =\langle z \rangle$ with the complete absence of fluctuations ($\delta\tilde k_i=0$ and $\xi_i=0$).
In contrast, for a sparse network characterized by a finite $\langle k \rangle$ such as the Erd\"os–R\'enyi (ER) network with an extremely low connection probability~\cite{network_ex1, network_ex2} or scale-free networks~\cite{network_ex3}, the fluctuations can be of the order $\mathcal{O}(1)$.
Consequently, these fluctuations may influence the dynamics significantly, leading to steady states distinct from those of the CG version.

Here, we focus our attention on a dense network (DN) characterized by the following conditions:
\begin{equation} \label{eq:degree}
\langle k \rangle \propto N^\alpha 
\,\,\, {\rm{and}} \,\,
\sqrt {\langle k^2 \rangle - \langle k \rangle^2 } \propto N^\beta \,
\end{equation}
with $0 <\alpha \le 1$ and $\beta<\alpha$.
The above power-law scaling of $\langle k \rangle$ can be found in the ER network with a finite connection probability~\cite{network_ex2} and scale-free-like DNs~\cite{network_ex3}.
It is then straightforward to estimate the order of local fluctuations as
\begin{align}
\label{eq:hi3}
\delta\tilde k_i \sim  \mathcal{O} \left( N^{-(\alpha-\beta)} \right)\quad \textrm{and}\quad
\xi_i \sim \mathcal{O}\left(N^{-\alpha/2}\right) \,,
\end{align}
where we estimate $\xi_i\sim \mathcal{O}(\sqrt{k_i}/\langle k\rangle)$ under the reasonable assumption that the phase factor fluctuation ($z_j-\langle z\rangle$) is almost Gaussian random with zero mean. In DNs, both local fluctuations vanish as $N\to \infty$, and then
the local field $ h_i$ in Eq.~\eqref{eq:hi2} converges to the global order parameter $\langle z\rangle$ in the ordered phase with
$\langle z\rangle \sim \mathcal{O}(1)$. Consequently,
the dynamic equation of motion, Eq.~\eqref{eq:eom_network2}, becomes identical to that for the CG as $N\to \infty$ and
thus the order parameter values
for DNs become identical to those for the CG. 
It is noteworthy that  in the incoherent (disordered) phase, where $\langle z\rangle$ also approaches to zero as $N\to \infty$,
$\langle z\rangle$ competes with the local fluctuation $\xi_i$ in Eq.~\eqref{eq:hi2}.
This suggests that
different types of finite-size effects in the CG, DN, and AN may yield distinct incoherent patterns,
while still adhering to the condition of the vanishing order parameter.

\section*{Annealed approximation on networks}

The analytical treatment of models on a network with quenched link disorder is typically challenging.
Instead, the AA is frequently employed due to its analytical tractability, while still
yielding results analogous to the original system.

In the AA, networks are substituted with annealed ones using the heterogeneous mean field approximation~\cite{network1, network2, hetero}, where the linking probability depends solely on the degrees of the two connecting nodes.
Within this framework, the adjacency matrix $a_{ij}$ is replaced by $a^{\rm A}_{ij}$ as
\begin{equation} \label{eq:ann}
a^{\rm A}_{ij} = \frac{k_i k_j} {N\langle k \rangle } \,,
\end{equation}
which represents the mean linking probability between nodes $i$ and $j$~\cite{hetero, annealed3}.
Then,
the annealed version of the local field $h^{\rm A}_i$ is given by
\begin{equation}
\label{eq:hi_a1}
h^{\rm A}_i = \frac{1}{\langle k \rangle} \sum_{j}
a^{\rm A}_{ij} z^{\rm A}_j
=\langle z^{\rm A} \rangle \left ( 1 + \delta\tilde k_i \right) + \xi^{\rm A}_i\,
\end{equation}
with phase factor $z^{\rm A}_j = e^{{\rm i} \phi_j^{\rm A}}$ satisfying the AA-applied dynamic equation.
The average phase factor fluctuation $\xi^{\rm A}_i$ in the AA is rather simplified as
\begin{equation}
\label{eq:xi_a1}
\xi^{\rm A}_i = \frac{1}{\langle k \rangle} \sum_{j}
a^{\rm A}_{ij}\left( z^{\rm A}_j -\langle z^{\rm A} \rangle \right)
= \frac{k_i}{\langle k\rangle} \left( \langle h^{\rm A} \rangle-\langle z^{\rm A}\rangle\right) \,.
\end{equation}
Note that the local property of $\xi^{\rm A}_i$ is limited to $k_i$ (no explicit connection information), leading to the
simple expression for $h^{\rm A}_i$ as
\begin{equation}
\label{eq:hi_a2}
h^{\rm A}_i =\frac{k_i}{\langle k\rangle}\langle h^{\rm A} \rangle \,,
\end{equation}
whose local property is also solely given by $k_i$, allowing the analysis of the dynamic equation~\eqref{eq:eom_network2} simpler.
The order of $\xi^{\rm A}_i$ is estimated similarly as
\begin{equation}
\label{eq:xi_a2}
\xi^{\rm A}_i = \left( 1 + \delta \tilde k_i \right ) \frac{1}{N } \sum_{j} \delta \tilde k_j z^{\rm A}_j \lesssim \mathcal{O}\left( N^{-(\alpha-\beta)}\right)\,,
\end{equation}
which also vanishes as $N\to \infty$ for DNs, as expected.
Nonetheless, the finite-size effects are different for the annealed and quenched networks,
which may exert a crucial influence on the determination of incoherent steady states.

\subsection*{Validity of the annealed approximation for the WS model}

For the WS model with identical oscillators ($\omega_i=\Omega$)~\cite{ws1, ws2}, the dynamic equation is given by
$\dot{\phi}_i = \Omega + J~{\rm {Im}}\left(h_i z_i^* \right)$. We can set $\Omega=0$ without loss of generality
via simple mapping of $\phi_i \rightarrow \phi_i+\Omega t$.
On the CG, the local field loses its locality completely as $h_i=\langle z\rangle$, thus we get
\begin{equation}
\label{eq:eom_identical_cg}
\dot{\phi}_i = J~{\rm {Im}}\left(\langle z\rangle z_i^* \right) \, .
\end{equation}
In the long time limit, there exist two types of stable fixed points $\{\phi_i\}$, satisfying (a) $\langle z\rangle=0$
for $J<0$ (see Methods) or (b) $\phi_i=\Phi$ for $J>0$ with the global angle $\Phi$ defined by $\langle z\rangle =|\langle z\rangle|e^{\rm{i}\Phi}$.
Consequently, we obtain
$|\langle z \rangle| =1$ for $J>0$ and $\langle z\rangle =0$ for $J<0$ with  $\{\phi_i \}$
satisfying discrete rotational symmetries ($\sum_{j=1}^N z_j=0$)~\cite{ws1, ws2}.
An initial condition selects one of these many incoherent ``regular'' symmetric steady states for $J<0$,
exact for any finite $N$.

In DNs, these incoherent regular fixed points are not stable any longer, due to strong fluctuations
in the local field $h_i$. Instead, $\phi_j$ becomes random with the uniform distribution
over $[0,2\pi]$ to make
$\langle z\rangle$ vanish in the $N\rightarrow\infty$ limit for $J<0$. This is confirmed by numerical simulations [not shown here].
In the AA of DNs, the local field fluctuations become weaker with a degree dependence only as
in Eq.~\eqref{eq:hi_a2}, leading to
\begin{equation}
\label{eq:eom_identical_aa}
\dot{\phi}_i^{\rm A} = \frac{J k_i}{\langle k\rangle}~{\rm {Im}}\left(\langle h^{\rm A}\rangle z_i^{\rm A *} \right) \, .
\end{equation}
Similar to the CG case, there are two types of stable fixed points, satisfying (a) $\langle h^{\rm A}\rangle=0$  for $J<0$ (see Methods) and
(b) $\phi_i^{{\rm A}}=\Psi^{\rm A}$ for $J>0$ with another global angle $\Psi^{\rm A}$ defined by $\langle h^{\rm A}\rangle =|\langle h^{\rm A}\rangle|e^{\rm{i}\Psi^{\rm A}}$.
Consequently, we obtain $|\langle z^{\rm A} \rangle| =1$ for $J>0$ and $\langle z^{\rm A}\rangle \rightarrow 0$ as $N\rightarrow\infty$
for $J<0$ with  $\{\phi_i^{{\rm A}} \}$
satisfying more complex discrete symmetries ($\sum_{j=1}^N k_jz_j^{{\rm A}}=0$). Again, the incoherent steady state
depends on the initial condition.

The incoherent steady state patterns are all different for the WS model on the CG, DN, and AN. It implies that
the AA fails to predict the stable steady-state solutions of Eq.~(\ref{eq:eom_network}) for DN systems, in particular when
the oscillators are identical. Certainly, this failure is not common in general DN systems.
An illustrative example is evident in the Kuramoto model with frequency disorder in $\omega_i$.
In its incoherent phase in DNs, oscillators undergo rotational motion with their respective natural frequencies, i.e.~$\dot{\phi}_i = \omega_i$ in the long time limit. The order parameter $\langle z \rangle$  as well as the local field $h_i$ vanish as $N \to \infty$, identical to the behavior observed in the CG. The AA-applied systems are expected to yield the same rotating behavior, thereby
resulting in identical incoherent
states on the CG, DN, and AN. This is not much surprising, as the additional frequency disorder is of the order
$\mathcal{O}(1)$, dominant over local fluctuations that diminish in the $N\rightarrow\infty$ limit for DN systems.
Temporal fluctuations such as a thermal noise also destroy the intricate structure of incoherent regular steady states in the AA (and also in the CG), and then the incoherent steady state pattern becomes fully random, akin to the incoherent phase in DN systems.

An intriguing question arises as to the nature of additional disorder that maintains a distinction in
incoherent steady-state patterns between the AA-applied and original system. Subsequently, we investigate
the inclusion of a quenched coupling disorder residing in the links of DNs,
where the associated fluctuations are expected to be of a similar order to those of the quenched link disorder.

\section*{Oscillators with competing interactions}

We consider a generalized Kuramoto model with coupling disorder, governed by the
dynamic equation as
\begin{equation}
\label{eq:eom_osc_glass}
\dot{\phi}_i = \omega_i + \frac{1}{\langle k\rangle}\sum_{j=1}^N J_{ij} a_{ij}\sin\left( \phi_j - \phi_i \right)\, ,
\end{equation}
where $J_{ij}$ is a random coupling strength between oscillators $i$ and $j$.
For simplicity, we consider the CG case ($a_{ij}=1$, thus $\langle k\rangle=N$).
This model has been introduced and studied in the context of ``oscillator glass'' by Daido ~\cite{daido1, daido2}.

In the case of identical oscillators ($\omega_i=0$), this model  describes the zero-temperature Sherrington–Kirkpatrick model for
XY spin glass~\cite{inf_spin_glass1, inf_spin_glass2, inf_spin_glass3}. Recently, Hong and Martens~\cite{hong_first_order}
also considered this model with a probability distribution $P[J]$ for $J_{ij}$ that has two $\delta$-peaks such as
\begin{equation}
P[J]=p \delta (J-J_+) + (1-p) \delta (J-J_-) \, ,
\end{equation}
with $J_+>0$ (attractive) and $J_-<0$ (repulsive). These competitive interactions induce frustration between oscillators
and the first-order phase transition occurs between the fully ordered and disordered phases
when the mean value of coupling strengths ($pJ_+ + (1-p) J_-$) changes its sign, i.e.~
at $p=p_c=(-J_-)/(J_+-J_-)$~\cite{hong_first_order}.

It is now worth noting that the random interaction $J_{ij}$ in the so-called ``two-peak" model with identical oscillators can be expressed using the adjacency matrix $b_{ij}$ of a random network, as follows:
\begin{equation}
\label{eq:eom_model}
\dot{\phi}_i =  \frac{1}{N} \sum_{j=1}^N \left[ J_- + \left( J_+ - J_- \right) b_{ij} \right] \sin \left( \phi_j - \phi_i \right)\, ,
\end{equation}
where $b_{ij}=1$ represents a positive (attractive) link with $J_+$, while $b_{ij}=0$ represents a negative (repulsive) link with $J_-$.
As the positive and negative links are randomly distributed, the {\em positive} degree $k_i$ (the number of positive links stemming from node $i$) satisfies the binomial distribution with mean $\langle k \rangle =Np$ and variance $\langle k^2 \rangle - \langle k \rangle^2=Np(1-p)$ for large systems. Thus, the $\{b_{ij}\}$ network is dense with the exponents $\alpha=1$ and $\beta=1/2$ defined in Eq.~\eqref{eq:degree}. This model can be also regarded as a competition between the CG model with the negative coupling constant $J_-$ $(<0)$
and the DN model ($ b_{ij}$) with the positive one $p(J_+-J_-)$ $(>0)$. For more general case with the underlying DN ($a_{ij}$)
and multi-peak distributions of $P[J]$ in Eq.~\eqref{eq:eom_osc_glass}, we expect the coupled systems on multiple DNs like a hypernetwork ~\cite{ref:hypernetwork}, which are left for future study.

Eq.~\eqref{eq:eom_model} is rewritten in a more illustrative form as
\begin{equation}
\label{eq:eom_model_q}
\dot{\phi}_i =  \left(J_+ - J_-\right) {\rm Im} \left(q_i z^*_i  \right)\,,
\end{equation}
where the {\em corresponding} local field $q_i$  reads
\begin{equation}
\label{eq:q1}
q_i =  \frac{\langle k \rangle }{N}\, h_i - \Delta  \langle z \rangle \,,
\end{equation}
with $\Delta \equiv (-J_-)/\left(J_+ - J_-\right)=p_c$ $(0<\Delta<1)$.
Using  Eq.~\eqref{eq:hi2}, we get
\begin{equation}
\label{eq:q2}
q_i = \langle z \rangle \left( p-\Delta + p \delta \tilde k_i \right )  +  p \xi_i \,,
\end{equation}
with $\xi_i =\sum_{j} b_{ij}  \left(z_j - \langle z \rangle \right)/{\langle k \rangle }$.

\begin{figure}[ht!]
\includegraphics[width=\linewidth]{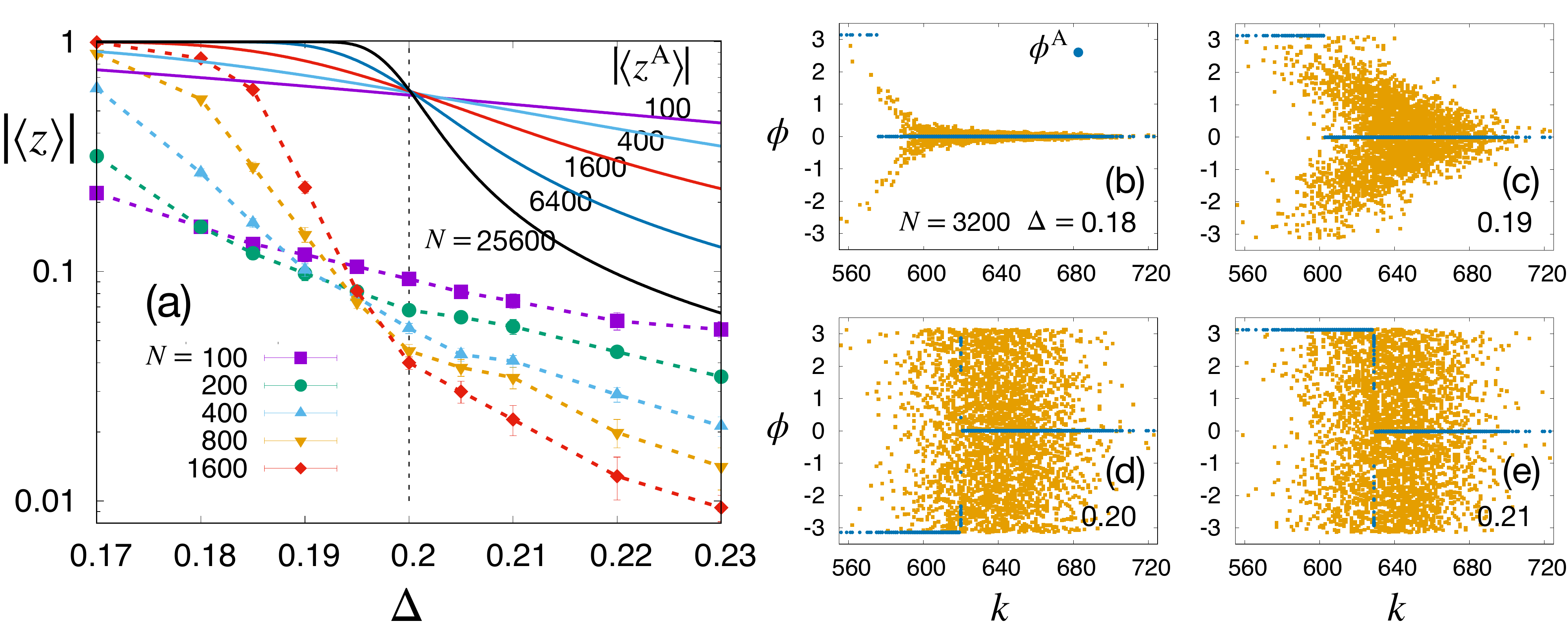}
\caption{
The original quenched system of the two-peak model versus
the annealed-network version.
Data points were obtained from numerical integrations of Eq.~\eqref{eq:eom_model} with the quenched network $b_{ij}$ and its annealed
version for $p=0.2$ and $J_+=1$.
(a) The order parameter as a function of $\Delta$ plotted on a semi-log scale for various values of $N$.
Solid symbols denote $|\langle z \rangle|$ for the original system and the dashed lines are guides to the eyes.
Solid lines represent the analytic solutions of $|\langle z^{\rm A} \rangle|$ obtained by the self-consistency equation
\eqref{eq:self1} for annealed networks.
Both the analytic and numerical results suggest a discontinuous transition
at $\Delta=p$ as $N\rightarrow\infty$.
The order parameter $|\langle z \rangle|$ at the transition seems to approach zero, while
its annealed version $|\langle z^{\rm A} \rangle|$ converges to a nontrivial value.
(b)-(e) Snapshots of phase angles $\phi$ and $\phi^{\rm A}$ versus degree $k$ in a single network realization of $N=3200$ near the steady state
for various values of $\Delta$. Blue dots represent $\phi^{\rm A}_i$ and yellow ones represent $\phi_i$.
The phase angle segregation in terms of degrees emerges in the annealed system, while it does not in the quenched case.
Note that the mean degree is given by $\langle k \rangle =Np= 640$.
}
\label{fig1}
\end{figure}

As discussed previously, both local fluctuations, $\delta \tilde k_i$ and $\xi_i$, in the DN diminish as $N\rightarrow \infty$, thus the local field $q_i$
approaches $\langle z\rangle(p-\Delta)$. Consequently, the dynamic equation becomes identical to Eq.~\eqref{eq:eom_identical_cg}
with substitution of $J$ by $(J_+-J_-)(p-\Delta)=pJ_++(1-p)J_-$ (mean coupling strength). Therefore, we expect the first-order phase transition at $p=\Delta$
from the disordered to the fully ordered phase. As the $\{b_{ij}\}$ network is dense (not CG for $p< 1$), there are finite-size fluctuations
which make the incoherent steady state uniformly random for $p\leq\Delta$ (numerically confirmed later, shown in Fig.~\ref{fig1}).

The application of the AA as in Eq.~\eqref{eq:ann} yields
\begin{equation}
\label{eq:eom_model_ann}
\dot{\phi}_i^{\rm A} =  \left(J_+ - J_-\right) {\rm Im} \left(q_i^{\rm A} z^{{\rm A}*}_i  \right)\,,
\end{equation}
with the annealed local field $q_i^{\rm A}$ given by
\begin{equation}
\label{eq:qa}
q^{\rm A}_i = \frac{k_i}{N } \langle h^{\rm A}\rangle -\Delta \langle z^{\rm A}\rangle\,,
\end{equation}
where Eq.~\eqref{eq:hi_a2} is used.
Similar to the previous cases without coupling disorder shown in Eqs.~\eqref{eq:eom_identical_cg} and \eqref{eq:eom_identical_aa},
there exist incoherent regular fixed points satisfying both $\langle h^{\rm A}\rangle=\langle z^{\rm A}\rangle =0$ simultaneously.
However, these regular fixed points are proven to be always unstable with a distribution of $k_i$ (see Methods).

Instead, one of two ``ordered'' fixed points becomes stable
in the long-time limit, depending on the magnitude of degree $k_i$;
\begin{equation}
\phi^{\rm A}_i = \left \{
\begin{array}{cc}
 \Phi^{\rm A} &\,\,\, {\rm for}\,\,\, k_i > k^* \\
\Phi^{\rm A} +\pi &\,\,\, {\rm for}\,\,\, k_i < k^*
\end{array} \right. \,,
\end{equation}
with $\langle z^{\rm A}\rangle =|\langle z^{\rm A}\rangle|e^{\rm{i}\Phi^{\rm A}}$.
The stability condition for the fixed points requires the identical global phase angles, denoted as
$\Phi^{\rm A}=\Psi^{\rm A}$ where $ \langle h^{\rm A}\rangle =|\langle h^{\rm A}\rangle|e^{\rm{i}\Psi^{\rm A}}$.
Thus, the threshold value $k^*$ is determined by the equation ${k^*}|\langle h^{\rm A}\rangle|/N -\Delta |\langle z^{\rm A}\rangle| =0$
(see Eq.~\eqref{eq:qa}),
which can be solved in a self-consistent manner. Detailed derivations are given later.

The coexistence of two coherent (ordered) clusters with a phase difference of $\pi$
is observed across the entire parameter range of $(p,\Delta)$. In the disordered phase ($p<\Delta$),
the threshold value $k^* \simeq \langle k\rangle$, leading to identical cluster sizes
in the $N\rightarrow\infty$ limit. Thus, the order parameters approach zero; $\langle z^{\rm A}\rangle\simeq 0$ and
$\langle h^{\rm A}\rangle\simeq 0$. Conversely, in the ordered phase ($p>\Delta$), one cluster fully dominates over the other in the
$N\rightarrow\infty$ limit with $k^*\simeq (\Delta/p)\langle k\rangle$. This results in
$\langle z^{\rm A}\rangle\simeq  1$ and $\langle h^{\rm A}\rangle\simeq 1$.

It is evident that the incoherent steady state for the AA-applied system differs from the random incoherent one in the original system.
This incoherent steady-state pattern exhibits the simple $Z_2$ symmetry (the same number of oscillators with
$z_i^{\rm A} e^{-\rm{i}\Phi^{\rm A}}= 1$ or $-1$), resembling a regular symmetric pattern
found in the WS model on the CG in Eq.\eqref{eq:eom_identical_cg}.
However, the origin of this symmetry is obviously different. Furthermore, this symmetry is exact only in the $N\rightarrow\infty$ limit and
is also independent of initial conditions, while the
regular symmetric patterns in the previous models are exact for any finite $N$ and are dependent on initial conditions.

\subsection*{Numerical results}

We numerically solve the equation of motion, Eq.~\eqref{eq:eom_model}, employing the Heun's method with various network sizes and fixed $p=\langle k \rangle/N$=0.2.
Initial phase angles are randomly chosen within $[-0.005 \pi, 0.005 \pi ]$. 
Solid symbols in Fig.~\ref{fig1} (a) denote the values of $|\langle z \rangle|$, averaged over a period of
$t=5\times 10^4 \sim 10^5$, after discarding an initial transient period of the same duration and also averaged
over $10 \sim 100$ network realizations and initial conditions.
We also display the analytic solutions $|\langle z^{\rm A} \rangle|$ in Fig.~\ref{fig1} (a), obtained from a self-consistency equation Eq.~\eqref{eq:self1} for the annealed networks.
We find that both $|\langle z \rangle|$ and $|\langle z^{\rm A} \rangle|$ seem to show a discontinuous transition
from the fully ordered to the disordered phase at $\Delta = p$, as $|\langle z \rangle |$ tends to approach 1 (0) asymptotically for $\Delta <p$ ($\Delta >p$) for increasing $N$.
At $\Delta =p$, however, $|\langle z^{\rm A} \rangle|$ remains finite as $N \to \infty$, as demonstrated by the crossing of solid curves, while $|\langle z \rangle|$ approaches zero.
This observation may suggest distinct underlying mechanisms of the transitions between the annealed and quenched systems.

In (b)-(e) in Fig.~\ref{fig1}, phase angle snapshots are plotted as a function of degree for the quenched and annealed cases, denoted by $\phi$ and $\phi^{\rm A}$, respectively. Data are obtained numerically from Eq.~(\ref{eq:eom_model}) in a single network realization for $p=0.2$ and $N=3200$.
Starting from an initial condition described above, the data points are obtained at $t=10^5$ near the steady state.
For $\Delta <p$, we observe in Fig.~\ref{fig1} (b) and (c) that in the original quenched system
a single coherent cluster with $\phi\simeq 0$ is formed by oscillators with higher degrees (stronger interactions effectively) and scattered phase angles
for those with lower degrees seem to be due to finite-size effects. The annealed case shows two coherent clusters with
phase difference of $\pi$, but one cluster dominates over the other. The finite-size effects are much weaker in the annealed systems, as
their fluctuations should be much weaker than those for the quenched systems.

For $\Delta >p$ in Fig.~\ref{fig1} (e), the phase angles seem randomly distributed as expected in the quenched system.
Remarkable distinction is found in the annealed system, where a binary mixture of two coherent clusters with comparable sizes
emerges. The contributions of the two clusters to the order parameter $\langle z^{\rm A}\rangle$ seem to cancel out exactly in
the $N\rightarrow\infty$ limit. At the transition ($\Delta=p$), the balance of two cluster sizes are slightly broken, yielding
a nontrivial value of $|\langle z^{\rm A} \rangle| \simeq 0.61946$, which is consistent with the analytic result derived in the following.

Back to the scattered plots in Fig.~\ref{fig1}, we note that
the outcomes from the original quenched model (orange dots)
present notable disparities in comparison to those of the annealed model (blue dots),
particularly in the vicinity of the threshold value $k^*$.
This phenomenon can be understood by considering the following.
In the quenched system, the magnitude of noise fluctuations $\xi_i$ is of the order $\mathcal{O} (N^{-1/2})$.
This suggests that the local field  $q_i$ is predominantly influenced by these fluctuations,
especially near $k\simeq k^*$, where
the value of the annealed local field almost vanishes as $q_i^{\rm A}\simeq 0$.
In contrast, in the region where $|k_i-k^*|/N \gtrsim \mathcal{O}(N^{-1/2})$, the annealed local field
$q_i^{\rm A}$ becomes comparable to $\xi_i$, resulting in
a reduced disparity between the quenched and annealed results. This trend is
also observable in the scattered plots.


\subsection*{Analytic solutions of the annealed two-peak model}

We rewrite the dynamic equation \eqref{eq:eom_model_ann} in a convenient form as
\begin{equation}
\label{eq:eom_model_ann1}
\dot{\phi}_i^{\rm A} =  \left(J_+ - J_-\right)
\left[  A_i \cos \tilde{\phi}_i^{\rm A} - B_i \sin \tilde{\phi}_i^{\rm A} \right] \,,
\end{equation}
where
\begin{align}
\tilde{\phi}_i^{\rm A}=\phi_i^{\rm A} - \Phi^{\rm A} \,, \quad
A_i = \frac{k_i}{N} |\langle h^{\rm A}\rangle| \sin (\Psi^{\rm A}-\Phi^{\rm A}) \,, \quad
B_i =  \frac{k_i}{N} |\langle h^{\rm A}\rangle| \cos (\Psi^{\rm A}-\Phi^{\rm A}) -\Delta |\langle z^{\rm A}\rangle| \,,\label{eq:abi}
\end{align}
yielding the stable steady-state fixed points as
\begin{align}
\label{eq:ssi}
\sin \tilde{\phi}_i^{\rm A}= \frac{A_i}{\sqrt{A_i^2+B_i^2}} \,, \qquad \cos \tilde{\phi}_i^{\rm A}= \frac{B_i}{\sqrt{A_i^2+B_i^2}} \,,
\end{align}

Utilizing the definition of the global phase angle $\Phi^{\rm A}$, we obtain the expression as
$|\langle z^{\rm A}\rangle|=z^{\rm A} e^{-\rm{i}\Phi^{\rm A}}= \langle e^{\rm{i}\tilde{\phi}^{\rm A}}\rangle$, implying
$\langle \sin \tilde{\phi}^{\rm A} \rangle=0$ and $\langle \cos \tilde{\phi}^{\rm A} \rangle \ge 0$.
In addition, from Eqs.\eqref{eq:abi} and~\eqref{eq:ssi}, we observe that the signs of
the steady-state solution for $\sin \tilde{\phi}_i^{\rm A}$
should coincide with the sign of $\sin (\Psi^{\rm A}-\Phi^{\rm A})$, independent of $i$. Then,
the constraint of $\langle \sin \tilde{\phi}^{\rm A}\rangle =0$ demands
$\sin \tilde{\phi}_i^{\rm A}=0$ for all $i$'s, thereby $A_i=0$ and $\sin (\Psi^{\rm A}-\Phi^{\rm A})=0$.
Subsequently, it follows that $\cos \tilde{\phi}_i^{\rm A}=\pm 1$ and $\cos (\Psi^{\rm A}-\Phi^{\rm A})=\pm 1$ in the steady state.
With the selection of $\cos (\Psi^{\rm A}-\Phi^{\rm A})=-1$, $B_i$ is always negative, thus
Eq.~\eqref{eq:ssi} yields $\cos \tilde{\phi}_i^{\rm A}=- 1$ for any $i$. However, this contradicts the condition $\langle \cos \tilde{\phi}^{\rm A} \rangle \ge 0$, thereby
$\cos (\Psi^{\rm A}-\Phi^{\rm A})=+1$ (implying $\Psi^{\rm A}=\Phi^{\rm A}$) must be selected for stable fixed points.
In this case, the sign of $B_i$ changes as $k_i$ varies.
Consequently, $\cos \tilde{\phi}_i^{\rm A}=+1$ with $B_i>0$ for large $k_i$  and
 $\cos \tilde{\phi}_i^{\rm A}=-1$ with $B_i<0$ for small $k_i$, leading to
\begin{equation}
\label{eq:phi_a}
\tilde{\phi}_i^{\rm A} = \left \{ \begin{array}{cc}
 0 &\,\,\, {\rm for}\,\, k_i > k^* \\
\pi &\,\,\, {\rm for}\,\, k_i < k^*
\end{array} \right. \,,
\end{equation}
where $k^*$ is defined as
\begin{equation}
\label{eq:ks}
 \frac{k^*}{N}\, |\langle h^{\rm A} \rangle | - \Delta |\langle z^{\rm A} \rangle | =0\,.
\end{equation}

To determine the threshold value $k^*$, we need to solve Eq.~\eqref{eq:ks} in a self-consistent manner.
Using $|\langle z^{\rm A} \rangle| = \langle \cos \tilde{\phi}^{\rm A}\rangle$ and $|\langle h^{\rm A} \rangle| = \langle k \cos \tilde{\phi}^{\rm A}\rangle/\langle k \rangle $, the order parameters in the steady state are given as
\begin{align}
\label{eq:r&h1}
&|\langle z^{\rm A} \rangle |= -\frac{1}{N}\sum_{k_i <k^*} + \frac{1}{N}\sum_{k_i > k^*} = 1- 2\sum_{k<k^*} B(k,p)\,, \nonumber \\
&|\langle h^{\rm A} \rangle |= -\frac{1}{N \langle k \rangle }\sum_{k_i <k^*} k_i + \frac{1}{N \langle k \rangle }\sum_{k_i > k^*} k_i = 1- \frac{2}{\langle k \rangle}\sum_{k<k^*} k B(k,p)\,,
\end{align}
where the degree distribution $B(k,p)$ is the binomial distribution of degree $k$ for a given $p$.
For large $N$, it is well known that $B(k,p)$ can be approximated by the Gaussian distribution
of a continuous variable $k\in[-\infty, \infty]$ with mean $\langle k\rangle=Np$ and variance
$Np(1-p)$. In this continuum limit, Eq.~\eqref{eq:r&h1}
can be expressed in a simple form as
\begin{align}
\label{eq:r&h2}
&|\langle z^{\rm A} \rangle|= - {\rm erf} \left[ \frac{k^*-Np}{\sqrt{2N p(1-p)} }\right] \,, \nonumber\\
&|\langle h^{\rm A} \rangle|=  |\langle z^{\rm A} \rangle| + \sqrt{\frac{2 (1-p)}{\pi N p}} \exp \left[ -\frac{\left( k^*- Np \right)^2}{2 Np(1-p)} \right] \,,
\end{align}
where the error function is defined as ${\rm erf}[x] = \left(2/\sqrt{\pi} \right) \int_0^x dt\, \exp\left[-t^2 \right] $.
From the constraint $|\langle z^{\rm A} \rangle| \ge 0$, we note that $k^*$ cannot exceed $\langle k \rangle =Np$.

For convenience, we rewrite Eq.~\eqref{eq:r&h2} as
\begin{align}
\label{eq:r&h3}
&|\langle z^{\rm A} \rangle|= - {\rm erf} \left[X\right] \,, \nonumber\\
&|\langle h^{\rm A} \rangle|=  |\langle z^{\rm A} \rangle| + \frac{\varepsilon}{p\sqrt{\pi}} e^{-X^2} \,,
\end{align}
with
\begin{align}
X \equiv  \frac{k^*-Np}{N\varepsilon}\le 0 \, \qquad \textrm{and} \qquad  \varepsilon\equiv\sqrt{\frac{2p(1-p)}{N}}
\end{align}
By substituting Eq.\eqref{eq:r&h3} into Eq.~\eqref{eq:ks}, we obtain
\begin{equation}
\label{eq:self1}
{\rm erf}\left[X\right] e^{X^2} =  \frac{{\varepsilon}}{p\sqrt{\pi}}\left(\frac{k^*}{k^*-N\Delta}\right)
=\frac{{\varepsilon}}{p\sqrt{\pi}}\frac{\varepsilon X +p}{\varepsilon X+(p-\Delta)} \,.
\end{equation}
The solution for $X$ in Eq.~\eqref{eq:self1} provides the threshold value $k^*$ and also determine the
order parameter values in Eq.~\eqref{eq:r&h3} in the steady state.
These results are in very good agreement with the numerical outcomes obtained directly from the annealed equation ~\eqref{eq:eom_model_ann}
(not shown here).
For small $\varepsilon$ (large  $N$), the explicit expression for $X$ can be derived from Eq.~\eqref{eq:self1} as
\begin{equation}
\label{eq:sol_self}
X \approx \left\{
\begin{array}{cc}
-{(p-\Delta)}{\varepsilon}^{-1}\, \,\,\, & {\rm for}\,\,\Delta <p \,, \\
 X_c + \frac{1-2p}{p} \frac{X_c^2}{2X_c^2 +1}\varepsilon \, \,\,\, & {\rm for}\,\,\Delta =p \,, \\
 -\frac{1}{{2} (\Delta -p) }\varepsilon\, \,\,\, & {\rm for}\,\,\Delta >p \,,
\end{array} \right.
\end{equation}
with $X_c \approx -0.62006$, determined by solving the equation of ${\rm erf}\left[X_c\right] e^{X_c^2} X_c=  {1}/\sqrt{\pi}$
which is given by Eq.~\eqref{eq:self1} with $\Delta=p$ for small $\varepsilon$.

\begin{figure}[t]
\centering
\includegraphics[width=0.6\linewidth]{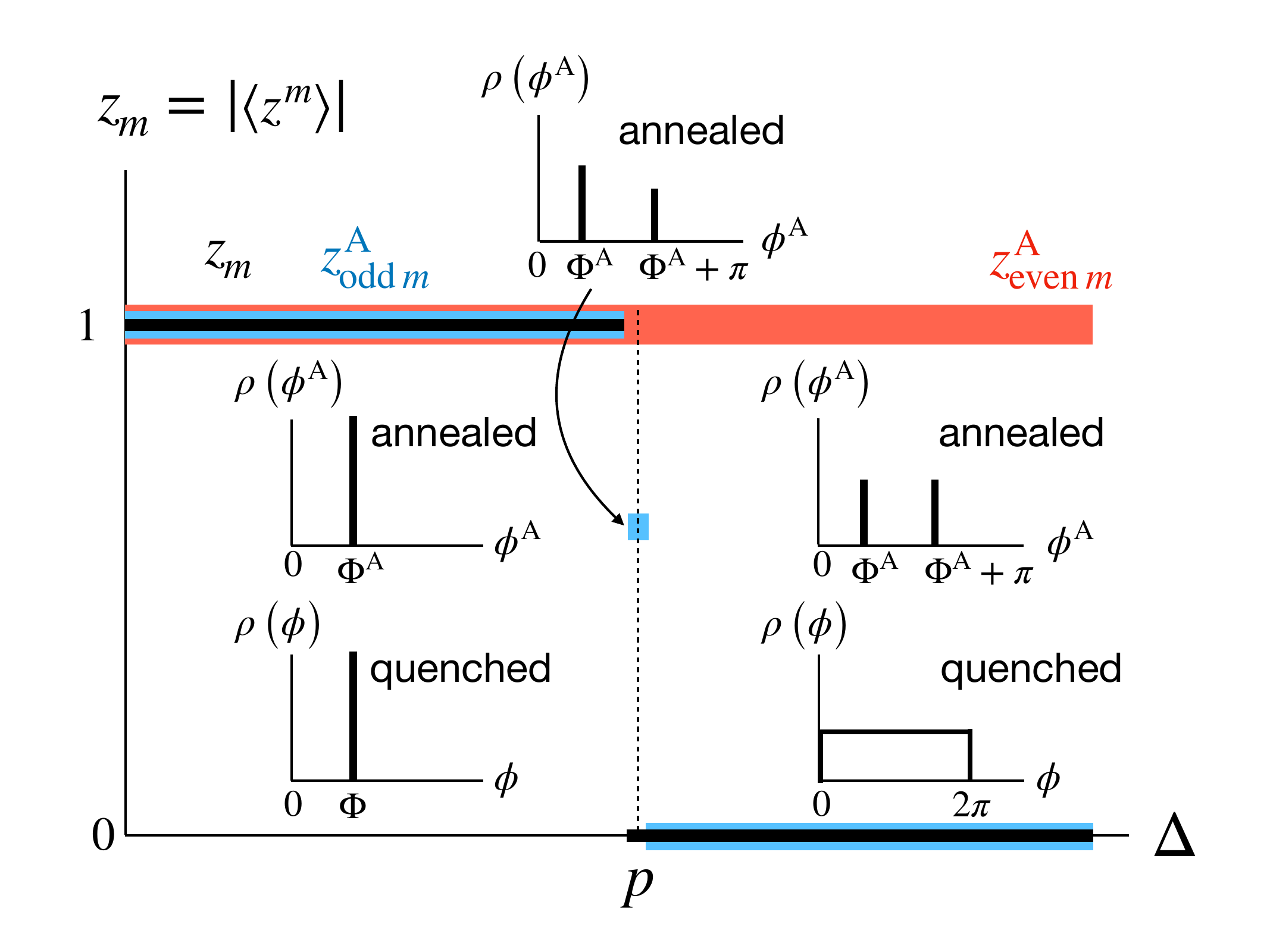}
\caption{Schematic diagram summarizing phase-angle distributions in annealed and quenched systems.
Black solid lines represent the generalized order parameters $z_m=|\langle z^m\rangle|$ for the quenched system with integer $m$.
Blue solid lines represent the generalized order parameter $z^{\rm A}_m$ for the annealed systems with odd $m$,
while red solid lines with even $m$.
}
\label{fig2}
\end{figure}

In the $N\rightarrow\infty$ ($\varepsilon\rightarrow 0$) limit,
the order parameters are calculated from Eq.~\eqref{eq:r&h3}:
$|\langle z^{\rm A} \rangle|=1$ ($X\rightarrow -\infty$), $|\langle z^{\rm A} \rangle|= -{\rm erf}[X_c] \approx 0.61946$, and  $|\langle z^{\rm A} \rangle|=0$ ($X\rightarrow 0$) for $\Delta<p$, $\Delta=p$, and $\Delta>p$, respectively, as seen in Fig~\ref{fig1} (a).
We also find $|\langle z^{\rm A} \rangle|\approx |\langle h^{\rm A} \rangle|$.
Eq.~(\ref{eq:sol_self}) is expressed in terms of $k^*$ as
\begin{equation}
\label{eq:ks_scaling}
 k^* -\langle k \rangle  \approx \left\{
\begin{array}{cc}
     -(p-\Delta)\, N \,\,\,  &   {\rm for}\,\,\Delta <p \,, \\
      X_c\sqrt{2p(1-p)}\, N^{1/2} \,\,\,  & {\rm for}\,\,\Delta =p \,, \\
     -p(1-p)/ (\Delta-p)\,\,\,  &  {\rm for}\,\,\Delta >p \,,
\end{array}
\right.
\end{equation}
yielding numerical values of $k^*$ as $576$, $608$, $620$, and $624$ for parameter values given in
Fig.~\ref{fig1} (b), (c), (d), and (e), respectively, that are in excellent accordance with simulation data.

The phase distribution $\rho(\phi^{\rm A})$ in the steady state is given by the combination of two $\delta$ peaks
from Eq.~\eqref{eq:phi_a} as
\begin{align}
\rho(\phi^{\rm A}) =   a\, \delta\left( \phi^{\rm A} -\Phi^{\rm A}\right) + b\, \delta \left( \phi^{\rm A} -(\Phi^{\rm A} +\pi )\right) \,,
\end{align}
where $a=\sum_{k>k^*} B(k,p)\approx \frac{1}{2} (1-{\rm erf}[X])$ and $b=1-a$. From Eqs.~\eqref{eq:sol_self} and \eqref{eq:ks_scaling},
we find $a=1$ ($b=0$) for $\Delta<p$ (single peak and fully synchronized) and $a=b=1/2$ for $\Delta>p$ (two symmetric peaks).
At $\Delta=p$, $a=\frac{1}{2} (1-{\rm erf}[X_c])\approx 0.80973$ ($b\approx 0.19027$) (two asymmetric peaks).
This is illustrated in Fig.~\ref{fig2}.

In large quenched systems, the numerical results suggest $\rho\left(\phi \right)=\delta\left(\phi- \Phi \right)$ for $\Delta <p$.
In contrast, for $\Delta \ge p$, it is observed that the phase angles are uniformly distributed, implying that $\rho(\phi)=1/({2\pi})$.
We can also consider the generalized order parameters defined as $z_m \equiv |\langle z^m \rangle| $ and $z^{\rm A}_m \equiv |\langle \left(z^{\rm A} \right)^m \rangle| $ with an arbitrary integer $m$.
We note that $z^{\rm A}_m$ for even $m$ does not distinguish the ordered and disordered phase ($z^{\rm A}_m=1$)
in the annealed systems, as two coherent clusters are synchronized with phase difference of $\pi$.
%

\section*{Summary}

In summary, we showed that the dynamic equations governing DN systems with quenched link disorder converge to those of the CG version in the thermodynamic limit, where the local fluctuations vanish entirely. Consequently, the AA-applied systems where fluctuations are
attenuated, exhibit the same dynamic behavior in the thermodynamic limit. However, a notable
discrepancy may arise in the incoherent (disordered) phase, where finite-size effects can become critical in
determining the steady-state pattern.

We illustrate our findings through two prototypical models of coupled oscillators;
the WS model for synchronization and the zero-temperature XY-like model with competing couplings.
In both cases, we analytically derive the incoherent patterns in the annealed systems, revealing stark differences from
those in the original quenched systems. These patterns in the annealed systems are regular and symmetric, in contrast to the random patterns in the quenched case. This suggests that caution should be given when applying the AA even to the DN systems, particularly
when examining incoherent steady-state patterns. 
We emphasize that our analysis based on the network topology and findings are not restricted to the oscillator or XY-spin systems. One may expect a similar discrepancy for the two-peak model with Ising spins because phase angles of oscillators with the AA-applied two competing interactions align along two branches, reminiscent of the Ising spins.
Furthermore, we point out the possibility of observing
these intriguing regular and symmetric incoherent patterns in temporal networks with sufficiently fast time scales,
which may underlie various biological and social systems.

\section*{Methods}
\subsection*{Stability analysis of incoherent regular solutions}

First, we perform a linear stability analysis of the incoherent regular fixed points for the WS model.
For the CG version with the dynamics governed by Eq.~\eqref{eq:eom_identical_cg},
a small perturbation  $\delta {{\phi}}_i$  around the fixed points ${{\phi}}_i^{\rm s}$ satisfying $\langle z\rangle=0$
evolves as
\begin{align}
\label{eq:lsa}
\delta \dot{{\phi}}_i=\sum_j S_{ij} \delta {{\phi}}_i \,,
\end{align}
with the stability matrix element $S_{ij}$ given by
\begin{align}
 S_{ij} = \frac{J}{N}\cos \left(\phi_j^{\rm s}- \phi_i^{\rm s}\right)  \,.
\end{align}
This stability matrix can be expressed as the sum of two rank-1 matrices as
$S_{ij} =S_{ij}^{(1)} + S_{ij}^{(2)}$ with $S_{ij}^{(1)}=(J/N) \cos (\phi_j^{\rm s})\cos (\phi_i^{\rm s})$ and
$S_{ij}^{(2)}=(J/N) \sin (\phi_j^{\rm s})\sin (\phi_i^{\rm s})$. It is trivial to show that any rank-1 matrix has
zero eigenvalues except for one eigenvalue given by its trace. 
For $J>0$, both rank-1 matrices are positive semi-definite with the non-zero eigenvalues of $(J/N)\sum_j \cos^2 (\phi_j^{\rm s})$ and $(J/N)\sum_j \sin^2 (\phi_j^{\rm s})$, respectively.
As the sum of two positive semi-definite matrices should be also positive semi-definite, it follows that 
the eigenvalues of the stability matrix are non-negative, leading to the conclusion that
all regular fixed points should be unstable for $J>0$. Consequently, the stabilization of the regular fixed points is only feasible for $J<0$.

This stability analysis can be also applied to the AN version with the dynamics governed by Eq.~\eqref{eq:eom_identical_aa},
leading to the stability matrix $S_{ij}^{\rm A}$  as
\begin{align}
 S_{ij}^{\rm A} = \frac{J}{N}\frac{k_i k_j}{\langle k\rangle^2} \cos \left(\phi_j^{\rm s}- \phi_i^{\rm s}\right)  \,.
\end{align}
It is clear that this stability matrix can be also decomposed into two rank-1 matrices, thus
the same conclusion as above can be drawn, as the link degree $k_i$ is always non-negative.

For the AA-applied model with coupling disorder, the dynamic equation is given by Eq.~\eqref{eq:eom_model_ann},
yielding the stability matrix as
\begin{align}
 S_{ij}^{\rm A} = \frac{{J_+-J_-}}{N}\left[\frac{p k_i k_j}{\langle k\rangle^2}-\Delta\right] \cos \left(\phi_j^{\rm s}- \phi_i^{\rm s}\right)  \,.
\end{align}
It should be noted that this matrix can be decomposed into 
two positive semi-definite rank-1 matrices and two negative semi-definite rank-1 matrices. 
It is straightforward but rather lengthy  to prove that the signs of the non-zero eigenvalues of the stability matrix 
align with those of the decomposed rank-1 matrices, if none of these rank-1 matrices is not a multiple of
another (a proof not included here). Given a distribution of $k_i$, it is evident that none of our four rank-1 matrices 
is proportional to any other. Consequently, the stability matrix invariably possesses two positive eigenvalues, thereby rendering all
incoherent regular solutions unstable. This conclusion is further substantiated by numerical results.

\section*{Acknowledgements}
We thank Su-Chan Park for valuable discussions. This research was supported by the NRF Grant No.~2022R1I1A1A01063166 (JU),
No.~2021R1A2B5B01001951 (HH), and
No.~2017R1D1A1B06035497 (HP), and
by the KIAS individual Grant No.~PG013604 (HP).

\section*{Author contributions statement}
All authors contributed to conducting this study and wrote the manuscript. 


\section*{Additional information}







\begin{thebibliography}{99}
\bibitem{network1} Dorogovtsev, S. N.,  Goltsev, A. V. \& Mendes, J. F. F. Critical phenomena in complex networks. {\it Rev. Mod. Phys.} {\bf 80}, 1275 (2008).
\bibitem{network2} Pastor-Satorras, R., Castellano, C., Van Mieghem, P. \& Vespignani, A. Epidemic processes in complex networks. {\it Rev. Mod. Phys.} {\bf 87}, 925 (2015).
\bibitem{hetero} Pastor-Satorras, R., Castellano, C., P. \& Vespignani, A. Epidemic spreading in scale-free networks. {\it Phys. Rev. Lett.} {\bf 86}, 3200 (2001).
\bibitem{annealed1} Ichinomiya, T. Frequency synchronization in a random oscillator network. {\it Phys. Rev. E} {\bf 70}, 026116 (2004).
\bibitem{annealed2} Lee, D.-S. Synchronization transition in scale-free networks: Clusters of synchrony. {\it Phys. Rev. E} {\bf 72}, 026208 (2005).
\bibitem{annealed3} Hong, H., Park, H. \& Tang. L.-H. Finite-size scaling of synchronized oscillation on complex networks. {\it Phys. Rev. E} {\bf 76}, 066104 (2007).
\bibitem{annealed4} Oh, E., Lee, D.-S., Kahng, B. \& Kim, D. Synchronization transition of heterogeneously coupled oscillators on scale-free networks. {\it Phys. Rev. E} {\bf 75}, 011104 (2007).
\bibitem{annealed5} Castellano, C. \& Pastor-Satorras, R. Routes to Thermodynamic Limit on Scale-Free Networks. {\it Phys. Rev. Lett.} {\bf 100}, 148701 (2008).
\bibitem{annealed8} Bogu\~{n}\'{a}, M., Castellano, C. \& Pastor-Satorras, R. Langevin approach for the dynamics of the contact process on annealed scale-free networks. {\it Phys. Rev. E} {\bf 79}, 036110 (2009).
\bibitem{annealed11} Yi, S., Um, J. \& Kahng, B. Extended mean-field approach for chimera states in random complex networks. {\it Chaos} {\bf 32}, 033108 (2022).
\bibitem{annealed6} Noh, J. D. \& Park, H. Critical behavior of the contact process in annealed scale-free networks. {\it Phys. Rev. E} {\bf 79}, 056115 (2009).
\bibitem{annealed7} Lee, S. H., Ha, M., Jeong, H., Noh, J. D. \& Park, H. Critical behavior of the Ising model in annealed scale-free networks. {\it Phys. Rev. E} {\bf 80}, 051127 (2009).
\bibitem{annealed9} Hong, H., Um, J. \&  Park, H. Link-disorder fluctuation effects on synchronization in random networks. {\it Phys. Rev. E} {\bf 87}, 042105 (2013).
\bibitem{annealed10} Um, J., Hong, H. \& Park, H. Nature of synchronization transitions in random networks of coupled oscillators. {\it Phys. Rev. E} {\bf 89}, 012810 (2014).
\bibitem{invalidity} Castellano, C. \& Pastor-Satorras, R. Non-Mean-Field Behavior of the Contact Process on Scale-Free Networks. {\it Phys. Rev. Lett.} {\bf 96}, 038701 (2006).
\bibitem{ref:KM} Kuramoto, Y. in {\it International Symposium on Mathematical Problems in Theoretical Physics} {\bf{30}}, 420 (Springer, New York, 1975); {\textit{Chemical oscillations, waves, and turbulence}} (Springer, Berlin, 1984).
\bibitem{review:KM} Acebr\'{o}n, J. A., Bonilla, L. L., Vicente, C. J. P., Ritort, F. \& Spigler, R. The Kuramoto model: A simple paradigm for synchronization phenomena. {\it Rev. Mod. Phys.} {\bf 77}, 137 (2005)
\bibitem{flat} Paz\'{o}, D. Thermodynamic limit of the first-order phase transition in the Kuramoto model. {\it Phys. Rev. E} {\bf 72}, 046211 (2005).
\bibitem{bimodal} Martens, E. A. {\it et al}. Exact results for the Kuramoto model with a bimodal frequency distribution. {\it Phys. Rev. E} {\bf 79}, 026204 (2009).
\bibitem{dense_network} Newman, M. {\it Networks} (Oxford University Press, Oxford, 2018).
\bibitem{dense_validity} Li, C., van de Bovenkamp, R. \& Van Mieghem, P. Susceptible-infected-susceptible model: A comparison of {\it N}-intertwined and heterogeneous mean-field approximations. {\it Phys. Rev. E} {\bf 86}, 026116 (2012).
\bibitem{adiabatic_elimination} Walls, D. \& Milburn, G. {\it{Quantum Optics}} (Springer-Verlag, Berlin, Heidelberg, 2008).
\bibitem{ws1} Watanabe, S. \& Strogatz, S. Integrability of a Globally Coupled Oscillator Array. {\it Phys. Rev. Lett.} {\bf 19}, 2391 (1992).
\bibitem{ws2} Watanabe, S. \& Strogatz, S. Constants of motion for superconducting Josephson arrays. {\it Physica D} {\bf 74}, 197 (1994).
\bibitem{network_ex1} P. Erd\"os, P. \& and R\'enyi, A. On random graphs I. {\it Publ. Math.} {\bf 6}, 290 (1959).
\bibitem{network_ex2} Newman, M., Strogatz, S. \& Watts, D. Random graphs with arbitrary degree distributions and their applications. {\it Phys. Rev. E} {\bf 64}, 026118 (2001).  
\bibitem{network_ex3} Ma, F., Wang, X., Wang, P. \& Luo, X. Dense networks with scale-free feature. {\it Phys. Rev. E} {\bf 101}, 052317 (2020).
\bibitem{daido1} Daido, H. Population Dynamics of Randomly Interacting Self-Oscillators. I: Tractable Models without Frustration. {\it Prog. Theor. Phys.} {\bf 77}, 622 (1987).
\bibitem{daido2} Daido, H. Quasientrainment and slow relaxation in a population of oscillators with random and frustrated interactions. {\it Phys. Rev. Lett.} {\bf 68}, 1073 (1992).
\bibitem{inf_spin_glass1} Sherrington, D. \& Kirkpatrick, S. Solvable Model of a Spin-Glass. {\it Phys. Rev. Lett.} {\bf 35}, 1792 (1975).
\bibitem{inf_spin_glass2} Kirkpatrick, S.\& Sherrington, D. Infinite-ranged models of spin-glasses. {\it Phys. Rev. B} {\bf 17}, 4384 (1978).
\bibitem{inf_spin_glass3} Billoire, A. Some Aspects of Infinite-Range Models of Spin Glasses:
Theory and Numerical Simulations in {\it Rugged Free Energy Landscapes} (ed. Janke, W.) 11-46 (Springer, Berlin, Heidelberg 2008).
\bibitem{hong_first_order} Hong, H. \& Martens, E. A. First-order like phase transition induced by quenched coupling disorder. {\it Chaos} {\bf 32}, 063125 (2022).
\bibitem{ref:hypernetwork} Sorrentino, F. Synchronization of hypernetworks of coupled dynamical systems. {\it New J. Phys.} {\bf 14}, 033035 (2012).
\end{thebibliography}
\end{document}